\documentclass[12pt,floats,showpacs,superscriptaddress,preprintnumbers,amsmath,amssymb,floatfix]{iopart}
\usepackage{graphicx}
\usepackage{dcolumn}
\usepackage{bm}

\begin{document}

\title[Spin polarized current and Andreev transmission]{Spin polarized current and Andreev transmission in planar
superconducting/ferromagnetic Nb/Ni junctions}
\author{E. M. Gonz\'alez$^1$, A.D. Folgueras$^{2,3}$, R. Escudero$^{1,*}$, J. Ferrer$^2$, F. Guinea$^3$ and J.L. Vicent$^1$}
\address{$^1$ Departamento de F{\'\i}sica de Materiales, Facultad de CC. F{\'\i}sicas,
Universidad Complutense, 28040 Madrid, Spain}
\address{$^2$ Departamento de F\'{\i}sica, Universidad de Oviedo, 33007 Oviedo, Spain}
\address{$^3$ Instituto de Ciencia de Materiales de Madrid, CSIC, Cantoblanco, 28049
Madrid, Spain}
\address{$^*$ On leave from Instituto de Investigaci\'on en Materiales,
Universidad Nacional Aut\'onoma de M\'exico, M\'exico D. F.}

\ead{ana@condmat.uniovi.es}

\begin{abstract}
We have measured the tunnelling current in Nb/Nb$_x$O$_y$/Ni planar tunnel
junctions at different temperatures. The junctions are in
the  intermediate transparency regime. We have extracted the
current polarization of the metal/ferromagnet junction without applying
a magnetic field. We have used a simple
theoretical model, that provides consistent fitting parameters for the
whole range of temperatures analyzed. We have also been able to gain insight into the
microscopic structure of the oxide barriers of our junctions.%
\end{abstract}

\maketitle

\section{Introduction}

Experiments with tunnel junctions using ferromagnetic metals\cite{Coleman}
have been an interesting topic since a long time. This subject has grown
again\cite{Akerman} because of the new field of spintronics where spin
dependent currents are an important requisite of many possible devices.\cite%
{Prinz, Boeck} This implies that control and measurements of spin polarized
currents are needed. Spin-polarized electron tunnelling\cite{Meservey} is a
key tool to measure the current polarization and to understand the physics
involved in these effects. Most of the recent experimental works have 
focused on the suppression of the Andreev reflection and have used the point 
contact geometry with a superconducting electrode\cite{Upadhyay, Soulen}. However
the local information extracted by point contact or scanning tunnelling
microscope techniques seems to be less suitable for devices than planar
tunneling junctions. In addition, the intrinsic difficulty of fabricating a
perfect uniform oxide layer can jeopardize the latter technique. Recently
Kim and Moodera\cite{Kim} have reported a large spin polarization of 0.25
from polycrystalline and epitaxial Ni (111) films using Meservey and
Tedrows's technique\cite{Meservey} and standard Al electrode and oxide
barriers, that allows for an almost ideal barrier behaviour.

In this work we show that the current polarization of ferromagnets can be
extracted without applying a magnetic field to the junction and using an
oxide barrier that is far from ideal. The experimental data are obtained for
Nb/Nb$_{x}$O$_{y}$/Ni planar tunnel junctions, where the barrier
is fabricated using the Nb native oxides. In this case, different Nb$_{x}$O$%
_{y}$ oxides are present in the barrier. This fact usually prevents the
analysis of the dV/dI characteristics in terms of perfect tunnelling. Indeed,
we show below that our junctions are neither in the tunnelling nor in the
transparent regime, but rather in a regime intermediate between both. We
shall argue that the current polarization can be obtained even in this
intermediate regime by use of a simple model. 

\section{Experimental method}

Nb(110) and Ni(111) films, grown by dc magnetron sputtering, were used as
electrodes. The structural characterization of these films was done by x-ray
diffraction (XRD) and atomic force microscopy (AFM), see for instance
Villegas \textit{et al}.\cite{PhysC369} Briefly, the junction fabrication
was as follows: First, a Nb thin film of 100 nm thickness was evaporated on
a Si substrate at room temperature. An Ar pressure of 1 mTorr was kept
during the deposition. Under these conditions, the roughness of the Nb film,
extracted from XRD and AFM, is less than 0.3 nm\cite{PhysC369} and
superconducting critical temperatures of 8.6 K are obtained. After this, the
film was chemically etched to make a strip of 1 mm width. A tunnel barrier
was prepared by oxidizing this Nb electrode in a saturated water vapor
atmosphere at room temperature.\cite{Halbritter} The thickness
of the oxide layer, extracted from the simulations performed with the
SUPREX program,\cite{Fullerton} is 2.5 nm. X-ray photoelectron spectroscopy 
analysis performed in these
oxidized films reveals that dielectric Nb$_{2}$O$_{5}$ is the main oxide
formed. There are also other oxides, such as metallic NbO, but in much less
amount. Taken into account Grundner and Halbritter studies,\cite{Halbritter}
Nb$_{2}$O$_{5}$ is the outermost oxide layer on Nb, whereas NbO is located
closer to the Nb film. The characterization by AFM reveals a RMS roughness
of around 0.7 nm. A detailed account of the structural and compositional
characterization of the barrier has been reported elsewhere.\cite{Elvira}

On top of this film (Nb with the oxide barrier), the second electrode of Ni
was deposited under the same conditions as Nb (up to 60 nm thickness) using
a mask to produce cross strips of 0.5 mm width, so that the overlap area $S$
of the two electrodes is 0.5 mm$^{2}$. 

Junctions fabricated using these materials and geometry, will not show a good
tunneling behaviour, neither a point-contact behaviour. As we will see, the
junctions will lie on the intermediate regime.

Perpendicular transport in tunnelling configuration was investigated by means
of characteristic dynamic resistance ($dV/dI$) versus voltage ($V$) using a
conventional bridge with the four-probe method and lock-in techniques. The
measured lock-in output voltage was calibrated in terms of resistance by
using a known standard resistor.

We have measured the conductance, defined as the inverse of the differential 
resistance $dV/dI$, of three different tunnel junctions. Moreover, junction 1 
(J1) has been measured and analyzed at temperature $T=1.52$ K, junction 2 (J2),
at temperatures $T=1.5$ and $T=3.945$ K, and junction 3 (J3), at temperatures 
$T=4.53$, $T=5.0$ and $T=5.39$ K. These measurements have allowed us to access and
assess the behavioiur and quality of our samples in the low (J1 and J2) and intermediate 
(J2 and J3) temperature range of this heterojunction.  Each data set presented
in this article has been normalized with respect to the background conductance $G_N$ of
the corresponding junction. 

\section{Theoretical Models}

\subsection{Introduction}

The conductance $G$ across a normal/superconducting junction may be expressed in
terms of the reflection probabilities $B$ of quasiparticles transversing the
junction, and of Andreev processes $A$ as 
\begin{equation}
G(V)\,=\frac{e^2}{h}\int\,d\epsilon\,\, (1+A-B)\,\,\frac{%
df(\epsilon-eV)}{d\epsilon}
\end{equation}
where $f(\epsilon)$ is the Fermi function; and $V$ is the applied voltage.

\begin{figure}[tbp]
\begin{center}
\includegraphics[width=8cm,height=7cm]{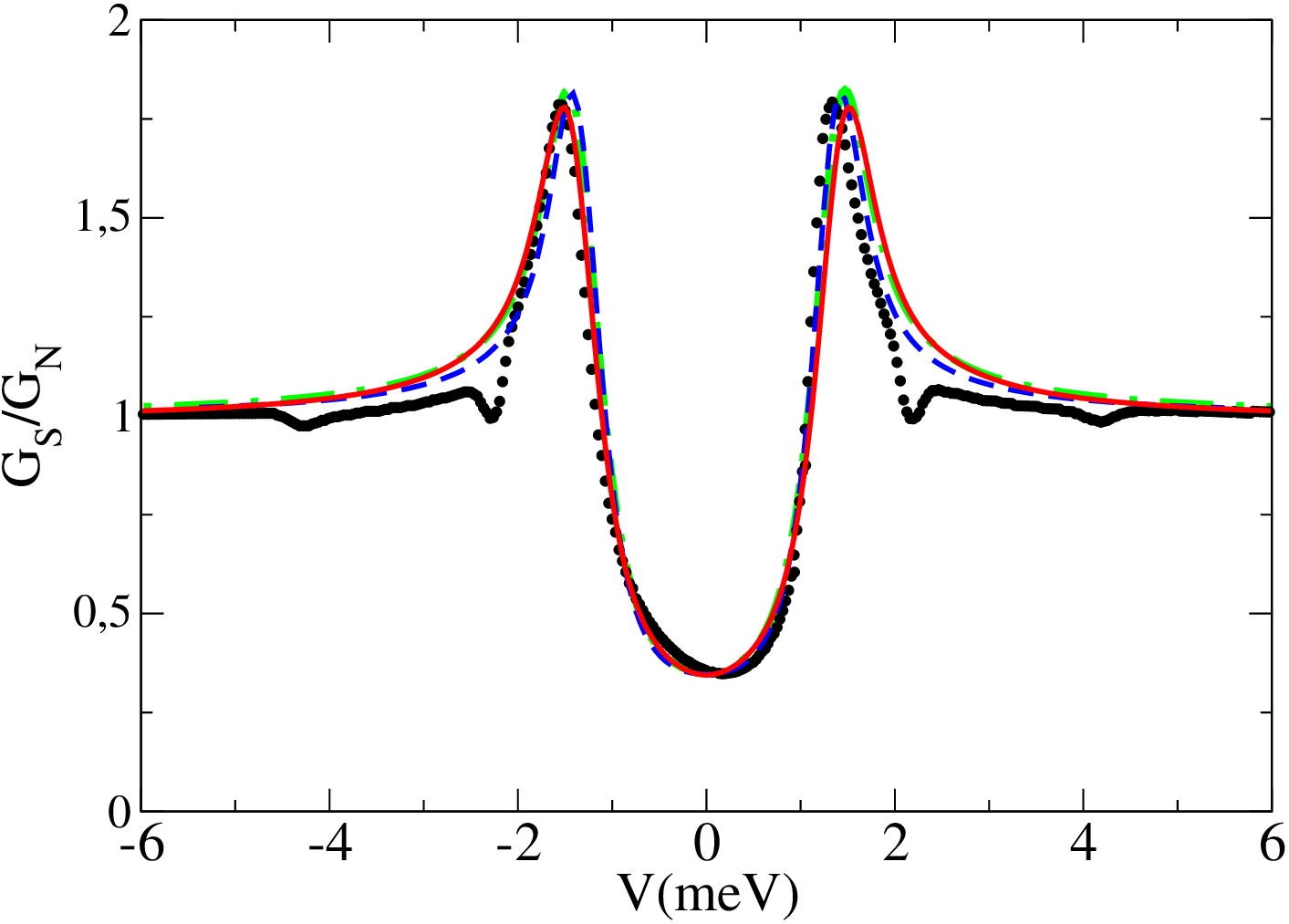}
\caption{Normalized tunneling conductance of junction 1 measured at $T=1.52$ K.
Black circles indicate experimental data; dashed-dotted green lines are fits to Model I,
dashed blue, fits to Model II, and solid red, fits to Model III.}
\end{center}
\label{fig1}
\end{figure}

\begin{figure}[t]
\begin{center}
\includegraphics[width=8cm,height=7cm]{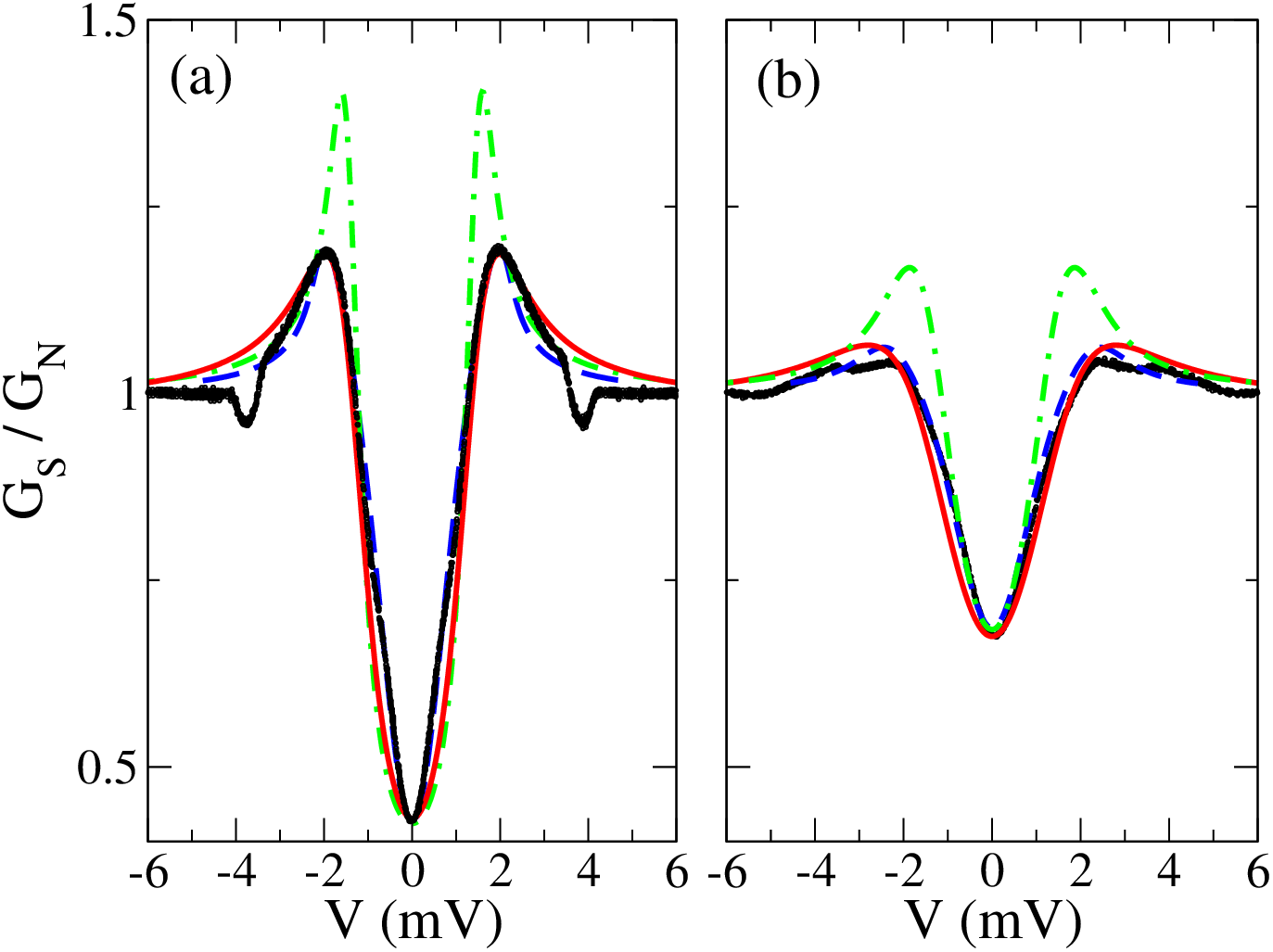}
\caption{Normalized tunneling conductance of junction 2
measured at $T=1.5$ K (a) and 3.945 K (b), together with a fit to 
Models I (dash-dotted green), II (dashed blue) and III (solid red).}
\end{center}
\label{fig2}
\end{figure}

Andreev reflection processes are proportional to the square of the
conventional transmission coefficient of the barrier, $T$, and therefore are
strongly suppressed for highly resistive barriers. Junctions with
transmission coefficients smaller than about 0.1 show small subgap
conductances, and can be classified as belonging to the tunneling regime.
Our experimental results, shown as black circles in ~\ref{fig1} and \ref{fig2}, exhibit a
significant conductance below the superconducting gap even at the lowest
temperatures. Therefore we expect that our effective oxide barriers should
be neither too high nor too thick, and we classify them as belonging to an
intermediate transparency regime.

While there are fairly complete descriptions of the transmission across 
ferromagnet/superconducting junctions\cite{mathias,kopu}, we have decided 
to describe it by three models 
that share the virtue that are conceptual and algebraically simple. These 
models are: i) The generalization of the Blonder-Tinkham-Klapwijk (BTK) model
\cite{BTK} to ferromagnetic electrodes proposed by Strijkers and 
coworkers;\cite{Ji} ii) a description of the effects of a finite current polarization
in terms of spin dependent transmission coefficients, as discussed by
P\'erez-Willard \textit{et al};\cite{Perez-Willard} and iii) a very simple
generalization of the BTK model to ferromagnetic electrodes with finite bulk
magnetization. 

We define the current polarization $P_c$ as the imbalance between the current
intensity of majority and minority carriers\cite{Soulen} in a
metal/ferromagnet junction, measured when the voltage tends to zero,

\begin{equation}
P_{c}=\frac{I_{\uparrow }-I_{\downarrow }}{I}\simeq\frac{G_{\uparrow
}-G_{\downarrow }}{G},
\end{equation}
where both spin channels contribute equally to the total current intensity
and conductance.

\subsection{Strijkers' model (Model I)}

Strijkers' model uses as adjustable parameters the current polarization $P_c$, the
height of barrier $Z$,\cite{Z} which is modeled by a delta function, and
the size of the superconducting gap at the interface $\Delta $.

The process of electron transfer across the junction is splitted into a fully
polarized channel, for which the Andreev reflection coefficient $A_{FP}$ is
zero, and 
\begin{eqnarray}
B_{FP}=\left\{%
\begin{array}{cc}
1 & \,\,\epsilon < \Delta \\ 
\frac{(u_0^2-v_0^2)^2Z^2(1+Z^2)}{(u_0^2+Z^2\,(u_0^2-v_0^2))^2} & \,\,
\epsilon > \Delta%
\end{array}
\right.
\end{eqnarray}
where $u_0$ and $v_0$ are the coherence coefficients of the superconducting
wave function; and a paramagnetic channel described by the coefficients $A_N$
and $B_N$ of the BTK model in its conventional form.\cite{BTK}

The total conductance is then written in terms of the conductance of the
fully polarized channel (${G_{P}}$) and the conductance of the paramagnetic
channel (${G_{N}}$): 
\begin{equation}
{{G\left( V\right) =\left( 1-P_{c}\right) G_{N}\left( V\right)
+P_{c}\,G_{P}\left( V\right) },}  \label{eq_conductance}
\end{equation}
This model interpolates between the paramagnetic case (BTK model), and the
half metal, where it predicts correctly that the amplitude for Andreev
reflection vanishes.

\subsection{Simple quasiclassical theory (Model II)}

We now use a simple model based on quasiclassical theory,\cite{Bel,Mor,Jiang,Serene} 
in which boundary conditions have been dumped into the spin-dependent transmission 
coefficients $T_\sigma$\cite{Perez-Willard}. The conductance
for each spin channel in this model is determined by 
\begin{equation}
G_\sigma(V)\,=\frac{e^2}{h}\,\int\,d\epsilon\,\,
(1+A_\sigma-B_\sigma)\,\,\frac{df(\epsilon-eV)}{d\epsilon}
\end{equation}
where the effective reflection coefficients depend now on spin. These can be
expressed in terms of the normal ${\cal G}$ and anomalous ${\cal F}$ components of the
Green's function, evaluated right at the superconducting side of the
interface, as follows 
\begin{eqnarray}
A_\sigma&=&T_\sigma\,\,T_{-\sigma}\,\,\left| \frac{{\cal F}}{%
1+r_\sigma\,r_{-\sigma}+(1-r_\sigma\,r_{-\sigma})\,\,{\cal G}}\right|^2 \nonumber
\\
B_\sigma&=&\left| \frac{r_\sigma+r_{-\sigma}+(r_\sigma-r_{-\sigma})\,\,{\cal G}} 
{1+r_\sigma\,r_{-\sigma}+(1-r_\sigma\,r_{-\sigma})\,\,{\cal G}}\right|^2,
\end{eqnarray}
where the reflection amplitude of the barrier satisfies the sum rule ${{r_\sigma^2+T_\sigma}=1}$.

The explicit functional form of ${\cal G}(\epsilon)$ and ${\cal F}(\epsilon)$ may be
obtained from the equation of motion of the quasiclassical Green function
that, for a bulk superconductor, reduces to 
\begin{eqnarray}
-\epsilon\,+\,i\,\Gamma\,{\cal G}&=&\Delta\,\frac{\cal G}{\cal F}\nonumber \\
{\cal G}^2-{\cal F}^2&=&1
\end{eqnarray}
where disorder and pair breaking effects, parametrized by the Dynes
parameter $\Gamma=\frac{\hbar}{2\,\tau}$,\cite{Dynes} are dealt within the
t-matrix approximation. Eq. (5) is used to fit the conductance data, using
the two transmissions $T_\sigma$, $\Gamma$ and $\Delta$ as adjustable
parameters.

\begin{figure}[tbp]
\begin{center}
\includegraphics[width=8cm,height=7cm]{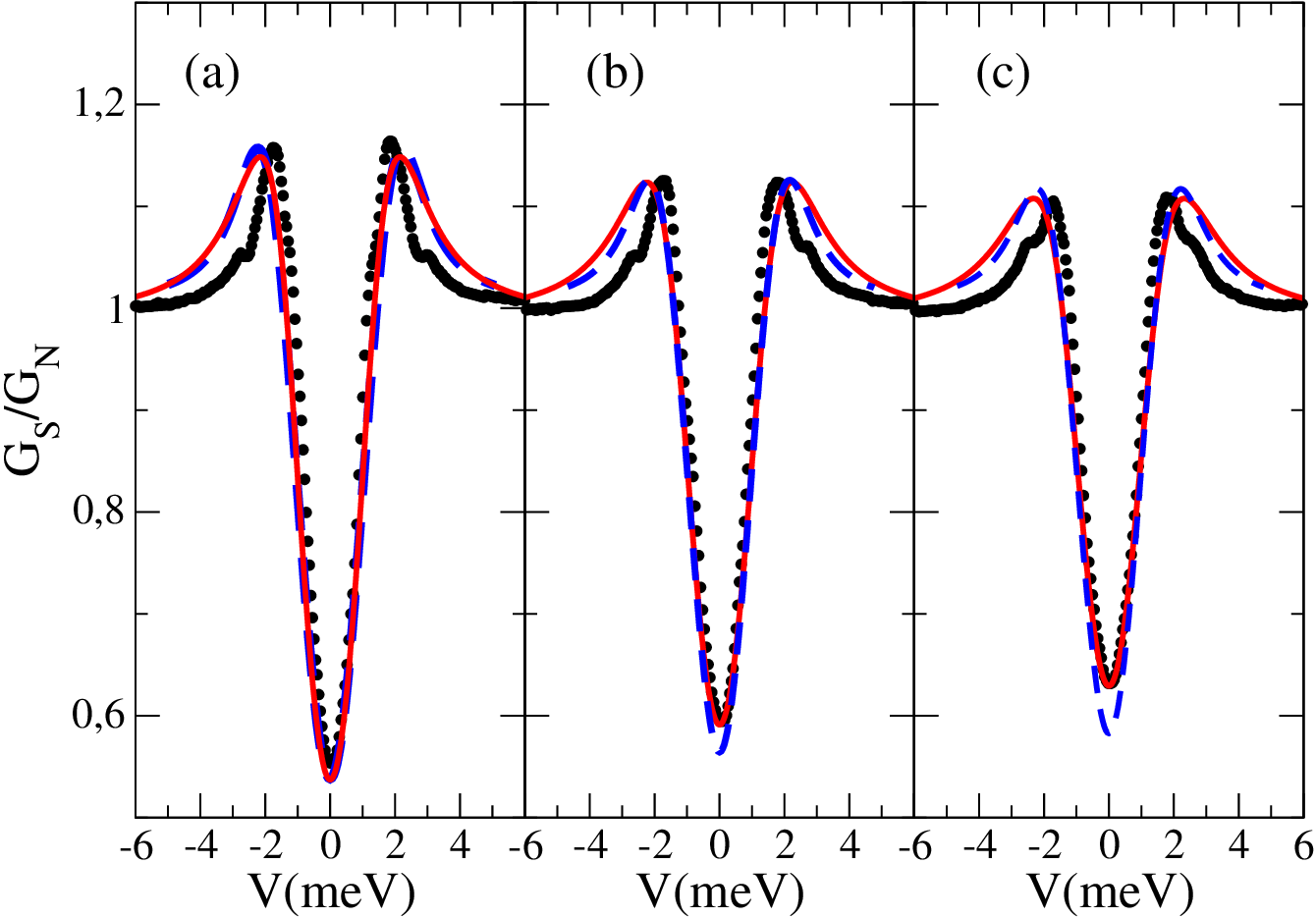}
\caption{Normalized tunneling conductance of junction 3
measured at $T=$ 4.53 K (a), 5.0 K (b) and 5.39 K (c), together with a fit to 
Models II (dashed blue) and III (solid red).}
\end{center}
\label{fig3}
\end{figure}

Eqs. (7) can also be solved when the gap $\Delta$ is zero. This gives
${\cal G}=1$ and ${\cal F}=0$, and corresponds the solution for the
normal state. Then, the conductance per spin chanel is, simply, 
$G_\sigma^0=T_\sigma$, and the current polarization is obtained from it as 
\begin{equation}
P_c=\frac{G_{\uparrow}^0-G_{\downarrow }^0}{G_\uparrow^0+G_\downarrow^0}=
\frac{T_\uparrow-T_\downarrow}{T_\uparrow+T_\downarrow}
\end{equation}

Perez-Willard et al.\cite{Perez-Willard} used this model to analyze Al/Co
point contacts, where the proximity effect may be discarded since the size
of the junction is negligible compared with the coherence length. They
indeed found close agreement with their experimental data for a wide range
of temperatures.

\begin{table}[tbp]
\begin{center}
\caption{Parameters used to fit junctions 1, 2, and 3 with Model II. Here 
$T_{av}=\frac{T_\uparrow + T_\downarrow}{2}$}
\begin{tabular}{cccccccc}
Junction & T (K) & $T_\uparrow$ & $T_\downarrow$ & $T_{av}$ & $P_c$ &$\Delta$ (meV) & $\Gamma$ (meV)\\
\hline
1 & 1.52 & 0.60 & 0.30 & 0.45 & 0.33 & 1.36 & 0.00\\
\hline
2 & 1.5  & 0.72 & 0.30 & 0.51 & 0.41 & 1.20 & 0.38\\
2 & 3.95 & 0.90 & 0.35 & 0.62 & 0.44 &  1.13 & 0.7\\
\hline
3 & 4.53 & 0.53 & 0.50 & 0.52 & 0.03 & 1.18 & 0.4 \\
3 & 5.00 & 0.43 & 0.40 & 0.42 & 0.04 & 1.10 & 0.5 \\
3 & 5.39 & 0.43 & 0.35 & 0.39 & 0.10 & 1.10 & 0.5 
\end{tabular}
\end{center}
\end{table}

\subsection{Generalization of BTK model for a ferromagnetic electrode (Model
III)}
We finally introduce ferromagnetism through an exchange splitting $J$ in one
of the electrodes. Hence, wave-vectors depend on spin as 
\begin{equation}
{{\hbar \,k_{\sigma}\,=m\,v_{F,\sigma}=\,(2\,m\,(E_{F}\,+\,\sigma \,J/2))^{1/2}}}
\end{equation} 

The barrier is modeled as a $\delta-$function of height Z. The normal 
$B_\sigma=|R_{B,\sigma}|^2$ and Andreev $A_\sigma=|R_{A,\sigma}|^2$ reflection 
probabilities, which depend on the spin flavour, can be calculated from 
\begin{eqnarray}
R_{A,\sigma} &=& \frac{ 2\,k_{-\sigma}\,q\,\Delta}{C+q^2\,R}\nonumber\\
R_{B,\sigma} &=& \frac{C-q^2\,R}{C+q^2\,R}
\end{eqnarray}
where the coefficient $C$ is equal to 
\begin{equation}
{(k_\sigma + k_{-\sigma} )\,q\,E +(k_\sigma k_{-\sigma}+i\,D\,(k_{\sigma}-k_{-\sigma})-D^2)\,R},
\end{equation}
$R=\sqrt{E^2-\Delta^2}$, the wave-vector $\hbar\,q$ is simply $\sqrt{2\,m\,(E_F+R)}$ and
$D= \hbar v_F Z$ is a parameter measuring the strength of the barrier.
We introduce disorder in a phenomenological
fashion,\cite{Dynes} by adding an imaginary part to the energy, $E=\epsilon - i \Gamma$.

The conductance per spin channel can be calculated in the same way as in the BTK
model using\cite{Gol} 
\begin{equation}
{{G_{\sigma }}=\frac{e^2}{h}\,\int \,{%
d\epsilon }\,\,(1+\frac{k_{-\sigma }}{k_{\sigma }}\,{A}_{\sigma }-{B}%
_{\sigma })\,\,\frac{df(\epsilon -eV)}{d\epsilon }}  \label{eq_conductance2}
\end{equation}

This model only depends on the four ratios $J/E_F, Z/E_F, \Gamma/E_F$ and $\Delta/E_F$, but
we prefer to fix $E_F$ instead of letting it dissapear by an adequate change of variables. We
therefore set $E_F=1.2$ eV, guided by our {\it Ab initio} simulation of Ni, performed with the
Molecular Dynamics suite SIESTA~\cite{bands}. We have also taken $J\approx0.8$ eV 
as representative of the spin-splitting of nickel along our experimental $\Gamma L$ direction.
The prefactor in front of the Andreev reflection amplitude is therefore set to  
$k_{-\sigma }/k_{\sigma}=1/\sqrt{2}$ from the outset.

The parameters $Z$, $\Gamma$ and $\Delta$ are on the contrary adjusted so
that formula (12) provide accurate fits to the conductance data. Once this is achieved, 
the current polarization at a given temperature is estimated by using again formula (12), 
but now with the gap set to zero.
\begin{equation}
P_c=\frac{G_{\uparrow}^0-G_{\downarrow}^0}{G_{\uparrow}^0+G_{\downarrow}^0}\approx
\frac{T_\uparrow-T_\downarrow}{T_\uparrow+T_\downarrow}
\end{equation}
where 
\begin{equation}
T_{\sigma} = 1 - B_{\sigma}^0
\end{equation}
are the transmission probabilities of each spin channel in the normal state.

\begin{figure}[tbp]
\begin{center}
\includegraphics[width=8cm,height=6cm]{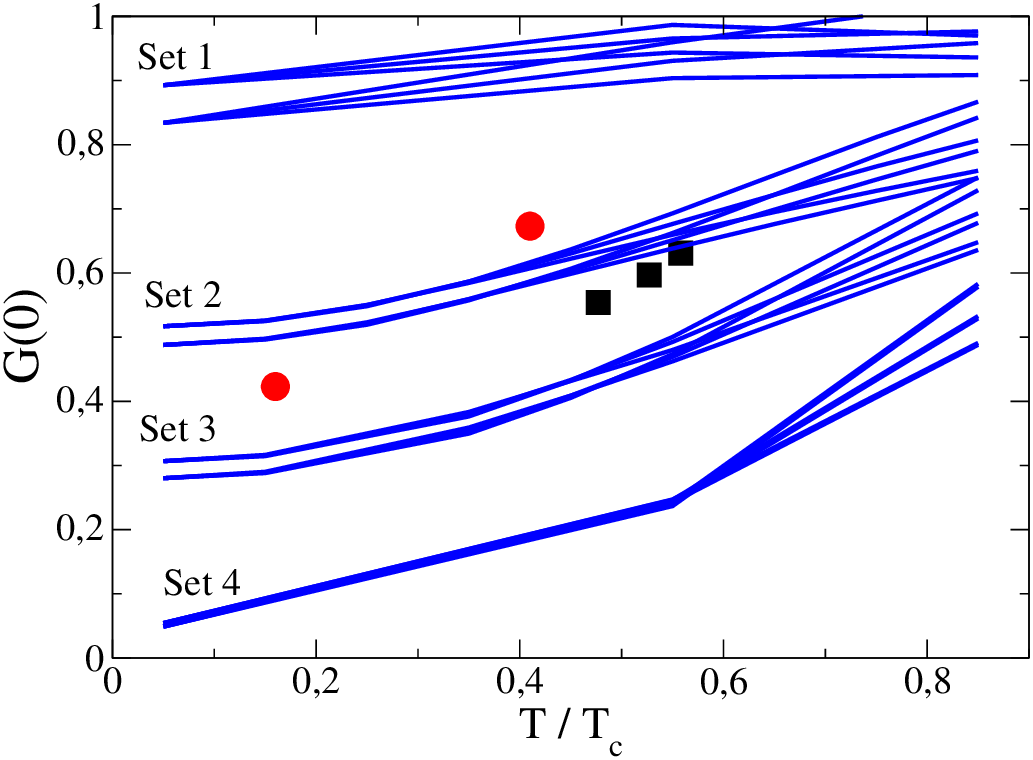}
\caption{Zero-voltage conductance $G(0)$ as a function of reduced temperature. Red circles
and black squares correspond to the data of J2 and J3, respectively. Sets 1, 2, 3 and 4 correspond
to $T_{av} = 0.15, 0.4, 0.55$ and 0.7, respectively. Each set has 6 lines, that correspond to the 
different combinations of $P_c=0.15$ and 0.35, and $\Gamma=0.1, 0.3$ and 0.5.}
\end{center}
\end{figure}

\section{Discussion}
\subsection{Comparison of theoretical models}
We plot the normalized conductance of J1, J2 and J3, as a function of 
voltage in Figs. 1, 2 and 3, respectively. J1, that has been measured at $T=1.5$ K, shows 
well developed coherence peaks. On the contrary, J2 and J3 do not display them even at low
temperatures. We have tried to fit the height and position of the coherence peaks, as well
as the height as shape of the low voltage conductance with the three models described above. 
We have not focused on the features that happen outside the gap in the curves, as physics in this region
is controlled by magnons and phonons, while inside the gap physics is controlled by Andreev processes, which determine the current polarization and do not interphere with phonon processes.

We have been able to fit J1 with Model I, but have failed to fit any of the conductance data of J2 and J3
with that model, that invariably gives too high coherence peaks, probably due to its 
oversimplified description of ferromagnetism and the neglect of disorder effects. 
This is explicitly shown in Figs. 1 and 2. 

We have been able to fit with Model II very accurately the conductance of junctions 1 and 2 for 
all temperatures. On the contrary, we have failed to fit the low voltage data 
of J3 at temperatures $T=5.0$ and 5.39 K, as shown in Fig. 3. 
Moreover, the fits to J2 and J3 provide transmission coefficients $T_\sigma$ that show a marked
dependence with temperature, as shown in Table I. Indeed, while $P_c$ 
remains essentially constant for a given junction, the spin-averaged transmission 
$T_{av}=(T_\uparrow+T_\downarrow)/2$ varies strongly 
with temperature. This is not physically correct, since the transmission coefficients should show
appreciable modifications only for temperature changes of the order of the bandwidth energy.
A closer look at the results presented in the table reveals that Model II gives a current
polarization that is unreasonably small for J3.

\begin{table}[tbp]
\begin{center}
\caption{Parameters used to fit junctions 1, 2 and 3 with Model III.} 
\begin{tabular}{cccccccc}
Junction & $T$ (K) &  Z & $T_\uparrow$ & $T_\downarrow$ & $P_c$ & $\Delta$ (meV) &$\Gamma$ (meV)\\
\hline
1 & 1.52 & 1.15 & 0.475 & 0.365 & 0.13 & 1.44 & 0.01 \\
\hline
2 & 1.5  & 3.02 & 0.115 & 0.078 & 0.19 & 1.44 & 0.57 \\
2 & 3.95 & 3.02 & 0.115 & 0.078 & 0.19 & 1.40 & 1.0 \\
\hline
3 & 4.53 & 2.00 & 0.228 & 0.161 & 0.17 & 1.35 & 0.35 \\
3 & 5.00 & 2.00 & 0.228 & 0.161 & 0.17 & 1.35 & 0.35 \\
3 & 5.39 & 2.00 & 0.228 & 0.161 & 0.17 & 1.35 & 0.35
\end{tabular}
\end{center}
\end{table}

It is also apparent from the table  that Model II provides values for the 
superconducting gap that are too small, and actually do not seem to follow 
the temperature dependence expected for a BCS superconductor. 
For instance, the model predicts a zero-temperature gap for J3, $\Delta_0 = 1.2$ meV. 
Using the measured critical temperature, we find that the ratio
$\frac{2\,\Delta_0}{k_B\,T_c}$ is equal to 3.25, which is much smaller than that of bulk Nb (3.8).

To understand better why the model fails
to fit the zero voltage conductance of J2 and J3, we plot in Fig. 4 $G(V=0)$ for several sets
of the parameters $T_{av}$, $P_c$ and $\Gamma$, that cover most of the parameter space. 
We have chosen for $T_{av}$ the values 0.1, 0.4, 0.55 and 0.7.
For each $T_{av}$, we have taken two representative values of $P_c=$ (0.15 and 0.35) and three different
values of $\Gamma$, (0.1, 0.3 and 0.5). The figure shows that the 24 curves cluster in four different sets,
according to the value of $T_{av}$. This implies that the value and temperature dependence of the zero-voltage
conductance is determined in this model essentially by the average transmision. The figure demonstrates in any
case that the experimental values of $G(0)$ for junctions J2 and J3 show a stronger dependence with 
temperature than the estimates provided by Model 2. We therefore believe that the model, while very 
appealing due to its simplicity, is actually too simple to describe the physical behavior of 
these junctions that belong to the intermediate transparency regime and
have been measured at low and intermediate temperatures. 

\begin{figure}[tbp]
\begin{center}
\includegraphics[width=8cm,height=6cm]{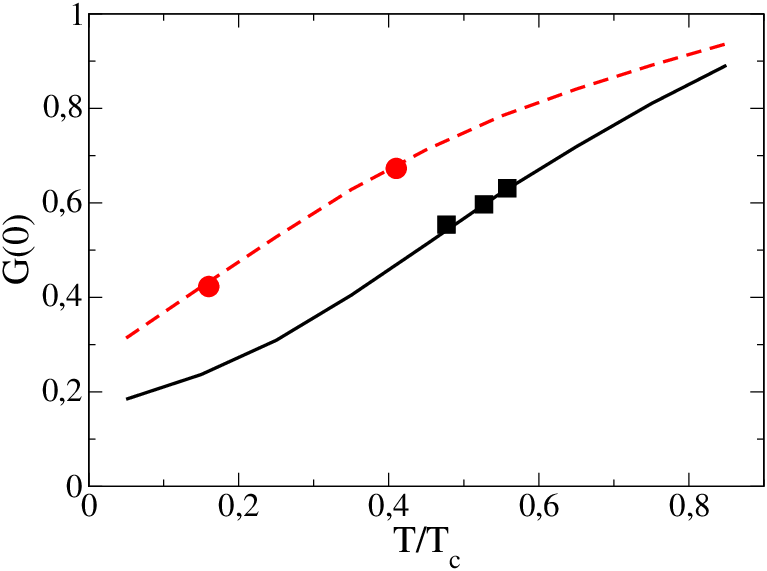}
\caption{Zero-voltage conductance $G(0)$ as a function of reduced temperature. Red circles
and black squares correspond to the experimentals data of J2 and J3, respectively. Dashed red and solid black
lines correspond to the values of $G(0)$ obtained with Model III, using the parameters shown in Table II.}
\end{center}
\end{figure}

We turn the attention now to our generalized BTK model (Model III). We note that the model is able to 
fit well the conductance data of 
junctions 1 and 2. In addition, it also provides a good fit to the data of J3, in contrast to
the quasiclassical model. More importantly, the model provides a temperature-independent barrier
height $Z$, that translates into temperature-independent transmission coefficients (see Table II). 
The values of Z so obtained allow us to clasify the junctions in an intermediate regime between 
tunnelling and point contact. Junction 2 is actually the closest to the tunneling regime, while 
Junction 1 is closest to the transparent regime. Junction 3, that we failed to fit with model II,
lies well within this intermediate regime. Fig. 5 clearly demonstrates that our model provides a temperature
dependence of $G(0)$ that fits well the experimental data.

Fig. 6 (a) shows the temperature dependence of the gap $\Delta$ obtained in the fits performed with 
Model III. We find values for the gap slightly smaller than those of bulk Nb, but consistent with the 
measured critical temperature and with the conventional temperature dependence of a BCS superconductor. 
The extrapolated zero-temperature gap, $\Delta_0=1.44$ meV, provides a superconducting ratio 
$\frac{2 \,\Delta_0}{k_B\,T_c}=3.90$ in close agreement with the ratio for Nb, 3.8. 

Fig. 6 (a) also shows the temperature dependence of the disorder parameter. We find that $\Gamma$ shows a smooth 
and linear dependence within the studied range of temperatures. The values of $\Gamma$ are large, as should
be expected, since our Nb samples are highly disordered.

We plot in Fig. 6 (b) the current polarization $P_c$ obtained using model
III, as a function of the height of the barrier Z. We find that $P_c$
increases with Z from 0.13 to 0.19, which shows that the current polarization 
increases with the tunelling quality of the junction.  We note that 
Soulen et al.~\cite{Soulen} have found polarizations of about 45\% using Ni/Nb junctions in a 
point contact geometry. More recently Kim and Moodera~\cite{Kim} have performed experiments for 
Ni/Al plannar tunnel junctions. They have found that $P_c$ grows from 11\% to 33\%
as the tunneling quality of the junctions increases. Our theoretical
calculations confirm that  
$P_c$ should increase with the strength or the thickness of the tunneling barrier.

The preceding analysis shows that the main advantage of model III
(BTK) over model II (adjustable spin dependent transmission coefficients)
is the stronger dependence on temperature of the conductance in model
III. The main difference between the two models lies in the description of
the ferromagnetic electrode. Model III uses two different Fermi surfaces,
one for each spin. Hence, the proximity effect is suppressed,
reducing the Andreev reflection at low temperatures.

\begin{figure}[tbp]
\begin{center}
\includegraphics[width=8cm,height=6cm]{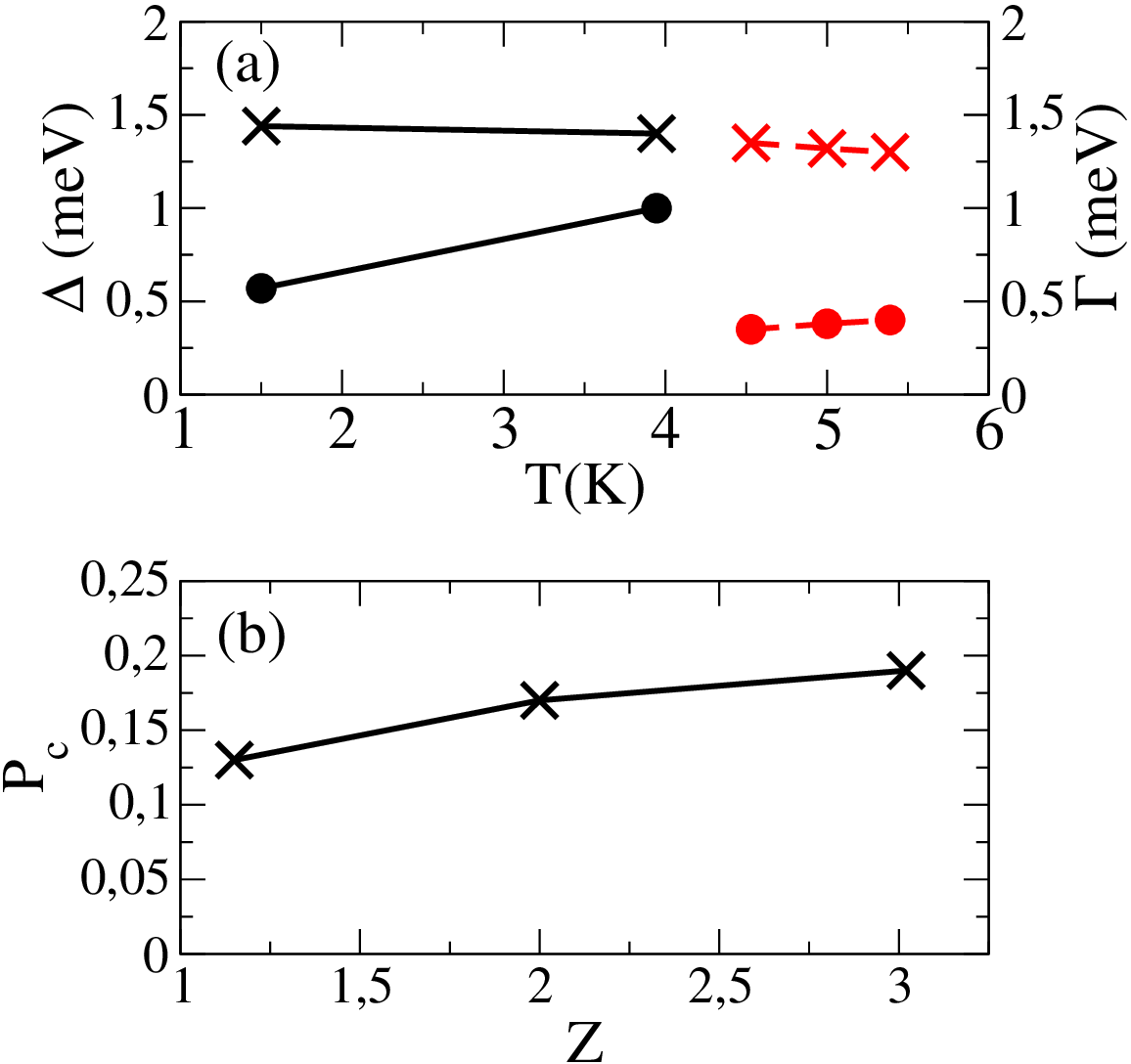}
\caption{(a) Values of the superconducting gap (crosses) and the Dynes
parameter (circles) as a function of temperature obtained from the fits of
our Generalized BTK model to the normalized conductance of junctions 2 (black)
and 3 (red). (b) Current polarization as a function of the height $Z$
of the delta-function barrier entering our Generalized BTK model for the
three junctions studied in this article. Lines have been added to aid the eye.}
\end{center}
\end{figure}

\section{Summary and conclusions}

We have fabricated and measured fairly large and far from ideal
superconducting/ferromegnetic tunel junctions. A simple theoretical model allows
us to extract tunnelling related parameters for such junctions. We have grown Nb/Nb$_x$O$_y$/Ni planar tunnel junctions, and
measured its conductance at different temperatures. We have found that
they belong to the regime of intermediate transparencies. We have been able
to fit the conductance curves with a simple generalization of the BTK 
model,\cite{BTK} that provides a sensible set of temperature-independent
transmission probabilities. Our calculations suggest that the results can
depend significantly on the description of the bulk ferromagnetic
electrode, as Andreev reflection depends both on the barrier transmission
coefficients and on the exchange field inside the ferromagnet. 
We have been able to give reasonable values 
of the current polarization without the need to apply a magnetic field. 
We have also studied the relationship between 
these two quantities, showing that the current polarization depends significantly
on the height of the barrier.
This simple generalization of the BTK Model to a superconducting/ferromagnetic junction
could be improved considering a 3D Model. However, we have shown that the simple
one dimensional model can fit the experimental data and also provides good estimates of the current polarization.

\section*{Acknowledgments}
We wish to thank very useful discussions with J. Moodera, M. Eschrig and M.
Fogelstrom. This work was supported by MEC (MAT2002-04543, MAT2002-12385E, 
MAT2002-0495-C02-01, BFM2003-03156 and AP2002-1383), and
CAM (GR/MAT/0617/2004). Additionally, E.M.G. acknowledges Spanish Ministerio
de Educaci\'on y Ciencia  for a Ram\'{o}n y Cajal contract, and R. Escudero
thanks Universidad Complutense and Ministerio de Educaci\'on y Ciencia for
a sabbatical professorship. A.D.-F. thanks Ministerio de Educaci\'{o}n y
Ciencia for a FPU grant (AP2002-1383).


\begin{thebibliography}{99}
\bibitem{Z} The height of the barrier Z in BTK's model is related to its
reflection coefficient in the normal state R, by $Z=R/\sqrt{1-R^2}$.
\end{thebibliography}
\end{document}